\documentclass[11pt]{article}
\pdfoutput=1
\usepackage{etex}

\usepackage[utf8]{inputenc}
\usepackage[T1]{fontenc}

\usepackage{xspace} 
\usepackage{numprint} 
\usepackage{mdframed}
\usepackage{balance}
\usepackage{makecell}
\usepackage{graphicx}
\usepackage{hyperref}
\usepackage[inline]{enumitem}
\usepackage{clipboard}
\usepackage[many]{tcolorbox}
\newtcolorbox{mybox}{enhanced,colframe=blue,
colback=yellow!4!white,
standard jigsaw,
  enlarge top by=.5cm,
  enlarge bottom by=.5cm,
}

\tcbuselibrary{skins}

\newclipboard{clipboard-botse}

\newcommand{\TODO}[1]{\textcolor{red}{todo: #1}}\newcommand\todo\TODO

\title{Explainable Software Bot Contributions:\\ Case Study of Automated Bug Fixes}
\author{Martin Monperrus\\ KTH Royal Institute of Technology\\\texttt{martin.monperrus@csc.kth.se}}
\date{}

\begin{document}

\maketitle

To appear in ``2019 IEEE/ACM 1st International Workshop on Bots in Software Engineering (BotSE)'' \medskip

\begin{abstract}
In a software project, esp. in open-source, a contribution is a valuable piece of work made to the project:
writing code, reporting bugs, translating, improving documentation, creating graphics, etc.
We are now at the beginning of an exciting era where software bots will make contributions that are of similar nature than those by humans.
 
Dry contributions, with no explanation, are often ignored or rejected, because the contribution is not understandable per se, because they are not put into a larger context, because they are not grounded on idioms shared by the core community of developers.

We have been operating a program repair bot called Repairnator for 2 years and noticed the problem of ``dry patches'': a patch that does not say which bug it fixes, or that does not explain the effects of the patch on the system.  We envision program repair systems that produce an ``explainable bug fix'': an integrated package of at least 1) a patch, 2) its explanation in natural or controlled language, and 3) a highlight of the behavioral difference with examples. 

In this paper, we generalize and suggest that software bot contributions must explainable, that they must be put into the context of the global software development conversation.
\end{abstract}

\section{Introduction}
The landscape of software bots is immense \cite{lebeuf:18}, and will slowly be explored by far and large by software engineering research.
In this paper, we focus on software bots that contribute to software projects, with the most noble sense of contribution: an act with an outcome that is considered concrete and valuable by the community.

The open-source world is deeply rooted in this notion of ``contribution'': developers are called ``contributors''. Indeed,  ``contributor'' is both a better and more general term than developer for the following reasons. First, it emphasizes on the role within the project (bringing something) as opposed to the nature of the task (programming). Second, it covers the wide range of activities required for a successful software project, way beyond programming: reporting bugs, translating, improving documentation, creating graphics are all essential, and all fall under the word ``contribution''.

Recently, we have explored one specific kind of contributions: bug fixes \cite{urli:hal-01691496,arXiv-1810.05806}. A bug fix is a small change to the code so that a specific case that was not well-handled becomes correctly considered. Technically, it is a patch, a  modification of a handful of source code lines in the program. The research area on automated program repair \cite{Monperrus2015} devises systems that automatically synthesize such patches. In the Repairnator project \cite{urli:hal-01691496,arXiv-1810.05806}, we went to the point of suggesting synthesized patches to real developers. Those suggestions were standard code changes on the collaborative development platform Github. In the rest of this paper, Repairnator is the name given to the program repair bot making those automated bug fixes.

A bug fixing suggestion on Github basically contains three parts: the source code patch itself, a title, and a textual message explaining the patch. The bug fixing proposal is called ``pull-request''. 
From a pure technical perspective, only the code matters. However, there are plenty of human activities happening around pull requests:
project developers triage them, integrators make code-review,
impacted users comment on them. 
For all those activities, the title and message of the pull requests are of utmost importance. Their clarity directly impact the speed of merging in the main code base. 

In the first phase of the Repairnator project \cite{urli:hal-01691496,arXiv-1810.05806}, we exclusively focused on the code part of the pull-request: Repairnator only created a source code patch, with no pull-request title and explanation, we simply used a generic title like ``Fix failing build'' and a short human-written message.
Now, we realize that bot-generated patches must be put into context, so as to smoothly integrate into the software development conversation. A program repair bot must not only synthesize a patch but also synthesize the explanation coming with it: Repairnator must create explainable patches.

This is related to the research on explainable artificial intelligence, or ``explainable AI'' for short \cite{gunning2017explainable}. Explainable AI refers to decision systems, stating that all decisions made by an algorithm must come with a rationale, an explanation of the reasons behind the decision. Explainable AI is a reaction to purely black-box decisions made, for instance, by a neural network. 

In this paper, we claim that contributions made by software bots must be explainable, contextualized. This is required for software bots to be successful, but more importantly, this is required to achieve a long-term smooth collaboration between humans and bots on software development.

To sum up, we argue in this paper that:

\begin{itemize}\itemsep=.1cm
\item Software bot contributions must be explainable.
\item Software bot contributions must be put in the context of a global development conversation.
\item Explainable contributions involve generation of natural language explanations and conversational features.
\item Program repair bots should produce explainable patches.
\end{itemize}

\begin{figure*}
\centering\includegraphics[width=.905\textwidth]{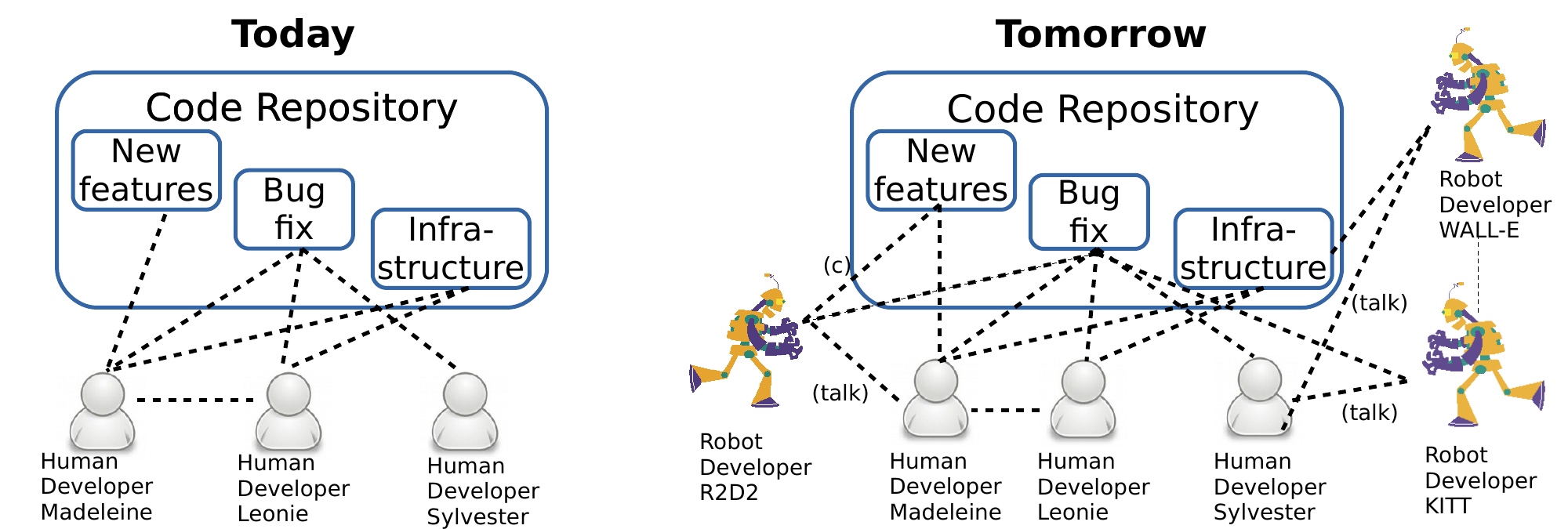}
\caption{One speculative future of software development, where robot developers and human developers smoothly cooperate.}
\label{fig-overview-vision}
\end{figure*}

Section \ref{sec:converstion} presents the software development conversation, Section \ref{sec:bots-as-communicating-agents} discusses why and how software bots must communicate.
Section \ref{sec:explainable-patch-suggestion} instantiates the concept in the realm of program repair bots.

\section{The Software Development Conversation}
\label{sec:converstion}

Software developers work together on so-called ``code repositories'' and software development is a highly collaborative activity. In small projects, 5-10 software engineers interact together on the same code base. In large projects, 1000+ engineers are working in a coordinated manner to write new features, to fix software bugs, to ensure security and performance, etc. In an active source code repository, changes are committed to the code base every hour, minute, if not every second for some very hot and large software packages.

\emph{All the interactions between developers is what forms the ``software development conversation''.}

\subsection{Nature of the Conversation}
The software development conversation involves exchanging source code of course, but not only.
When a developer proposes a change to the code, she has to explain to the other developers the intent and content of the change. Indeed, in mature projects with disciplined code review, \emph{all code modifications come with a proper explanation of the change in natural language}.
This concept of developers working and interacting together on the code repository is shown at the left-hand side of Figure \ref{fig-overview-vision}.

What is also depicted on Figure \ref{fig-overview-vision} is the variety of focus in the software development conversation. The developers may discuss about new features, about fixing bugs, etc. Depending on expertise and job title, developers may take only part to one specific conversation. On Figure \ref{fig-overview-vision}, developer Madeleine is the most senior engineer, taking part to all discussions in the project. Junior developer Sylvester tends to only discuss on bug reports and the corresponding fixing pull requests.

\subsection{Scale of the Conversation}
In a typical software repository of a standard project in industry, 50+ developers work together. In big open-source projects as well as in giant repositories from big tech companies, the number of developers involved in the same repository goes into the thousands and more.  For instance, the main repository of the Mozilla Firefox browser, \href{https://github.com/mozilla/gecko-dev}{gecko-dev}, has contributions from 3800+ persons. Table \ref{tab-extraordinary-repos} shows the scale of this massive collaboration for some of the biggest open-source repositories ever.

Notably, the software development conversation is able to transcend traditional organization boundaries: it works even when developers work from different companies, or even when they are only loosely coupled individuals as in the case of open-source.

\subsection{Channels}

The software development conversation happens in different channels. 

\emph{Oral channels} Historically, the software development conversation  happens in meetings, office chats, coffee breaks, phone calls. This remains largely true in traditional organizations.

\emph{Online channels} We have witnessed in the past decades the raise of decentralized development, with engineers scattered over offices, organizations and countries. In those contexts, a significant part of the software development conversation now takes place in online written channels:
mailing-lists, collaborative development platforms  (Github, Gitlab, etc), synchronous chats (IRC, Slack), online forums and Q\&A sites (Stackoverflow), etc.

\begin{mybox}
Source code contributions only represent a small part of the software development conversation. Most of the exchanges between developers are interactive, involving natural language. In the case of collaborative platforms such as Github, the bulk of the software development conversation happens as comments on issues and pull-requests.
Software bots will become new voices in this conversation.
\end{mybox}

\section{Software Bots as Communicating Agents}
\label{sec:bots-as-communicating-agents}

The software bot community now works on a different software development model, which is sketched at the right-hand side of Figure \ref{fig-overview-vision}.
Instead of only having human software developers working on a given code base, we will have code repositories on which both humans and bots would smoothly and effectively cooperate.
Here, cooperation means two things.
First that robots would be able to perform software development tasks traditionally done by humans: for instance a robot could be specialized in fixing bugs.
Second that robots and humans would communicate to each other to exchange information and take together informed decisions.

\begin{table}
\begin{tabular}{|l|r|r|}
Software & Commits & Contributors \\
\hline
\url{https://github.com/torvalds/linux} & 798710 &~~ n-a \\
\url{https://github.com/chromium/chromium} & 744581 &~~ n-a \\
\url{https://github.com/mozilla/gecko-dev} & 631802 & 3851 \\
\url{https://github.com/LibreOffice/core} & 433945 & 853 \\
\url{https://github.com/WebKit/webkit} & 208041 & ~~ n-a \\
\url{https://github.com/Homebrew/homebrew-core} &  135248 &  7310 \\
\url{https://github.com/NixOS/nixpkgs} & 166699 & 1935 \\
\url{https://github.com/odoo/odoo} & 122698 & 873 \\ 
\end{tabular}
\caption{Some of the biggest code repositories ever in the open-source world (data from Jan 2019)}
\label{tab-extraordinary-repos}
\end{table}

\subsection{Human Developers as Target}

Now, let us stress on the communication aspect of software bots. Software bots will not work alone, they will work together with human developers. As such, software bots must be able to communicate with human developers, using their language, given the human cognitive skills. Software bots will have to explain to other human developers what they are doing in each contribution, this is what can be called an \emph{``explainable contribution''}. 
Developers would likely ask clarification questions to bots, bots would answer and  this would form a proper engineering conversation.

\subsection{Software Bot Identity}
A good conversation is never anonymous. This holds for the conversation between human developers and software bots.

\emph{Name:} We believe it is important that software bots have a name and even their own character
In Figure \ref{fig-overview-vision} all bots are named.
Moreover, they are all named after a positive robot from popular culture.
Positive names and characters encourage developers to have a welcoming, forgiving attitude towards software bots.
We envision that software bots will be engineered with different characters: for example some would be very polite, others would be more direct à la Torvalds. 

\emph{Adaptation:} We envision that sophisticated software bots will be able to adapt the way they communicate to developers: for instance, they would not explain their contributions in the same way depending on whether they target a newcomer in the project or an expert guru developer. The tailoring of communication style may require project-specific training in a data-driven manner, based on the past recorded interactions  (or even developer specific training).

\emph{Diversity:} In all cases, we think that it is good that all bots are different, this difference is what makes  a development conversation rich and fun. Biodiversity is good, and similarly, we think that \emph{software bot diversity may be crucial for a bot-human community to thrive}.

\subsection{Contribution Format}
When human developers submit patches for review to other humans, it is known that the quality of the explanation coming with the patch is very important. The format depends on the practices and idiosyncrasies of each project. The patch explanation may be given in an email, as in the case of Linux, where the Linux Kernel Mailing List (aka LKML) plays a key role. The explanation may also be given as comment associated to a suggested modification, such as a pull request comment on the Github platform. A patch may be illustrated with some figure and  visualization of the change. A software bot must take into account the specificity of targeted platform and community.

\begin{mybox}
Many software bots will primarily produce contributions for humans. As such their contributions must match the human cognitive skills and must be tailored according to the culture of the targeted developer community. Humans prefer to interact with somebody, rather than with an anonymous thing, hence fun software bots will have a name and a specific character.
\end{mybox}

\section{Explainable Patch Suggestion}
\label{sec:explainable-patch-suggestion}

Now, we put the concept of ``explainable contribution'' in the context of a program repair bot \cite{Monperrus2015}. In this section, we explain that program repair bots such as Repairnator must explain the patches they propose.

We envision research on synthesizing explanations in order to accompany patches synthesized by program repair tools. 
The synthesized explanation would first be an elaborated formulation of the patch in natural language, or in controlled natural language.
Beyond that, the patch explanation would describe what the local and global effects on the computation are, with specific examples.
Finally, as done by human developers, the explanation could come with a justification of the design choices made among other viable alternatives.

\subsection{From Commit Summarization to Patch Explanation}
The goal of commit summarization is to reformulate a code change into natural language \cite{cortes2014automatically,JiangAM17,liu2018neural}. 
Commit summarization can be seen as both a broader task than patch suggestion (all changes can be summarized and not only patches), and smaller (only a few sentences, and even a single line in the context of extreme summarization are produced).
We envision experiments on using the state-of-the-art of commit summarization \cite{liu2018neural} on patches. It will be very interesting to see the quality of the synthesized summaries on patches, incl. on patches from program repair tools: do they capture the essence of the patch? is the patch explanation clear?

\emph{Visualization} A picture is worth 1000 words. Instead of text, the synthesized patch can be explained with a generated visualization. This idea can be explored based on research on commit, diff and pull-request visualization (e.g. \cite{DAmbros2010}).

\subsection{Automatic Highlighting of Behavioral Differences}

For humans to understand a behavioral difference, a good strategy is to give them an actual input and highlight the key difference in the output. 
There are works that try to identify valuable input points on which the behavior of two program versions differ \cite{marinescu2013katch,Shriver2017ESN}.
In a patch explanation, the selected input must satisfy two requirements: 
1) that input must be representative of the core behavioral difference introduced by the patch and 
2) it must be easily understandable by a human developer (simple literals, simple values).

The format of this behavioral difference  explanation is open. It may be a sentence, a code snippet, even a graphic. What is important is that it is both understandable and appealing to the developer. Importantly, the format must be set according to the best practices on communicating in code repositories (eg. communicating on Github).

\subsection{Conversational Program Repair Bots}

Finally, an initial explanation of a patch may not be sufficient for the humans developers to perfectly understand the patch. We imagine conversational systems for patch explanation: developers would be able to ask questions about the patch behavior, and the program repair bot would answer to those questions. Such a system can be data-driven, based on the analysis of the millions of similar conversations that have happened in open-source repositories.

\begin{mybox}
In the context of a program repair bot that produces bug fixes, an explainable bug fix means an integrated package: 1) a patch, 2) its explanation in natural language, and 3) a highlight of the behavioral difference with examples. 
The explanation might require a series of questions from the developers and answers from the bot, which requires advanced conversational features in the bot.
\end{mybox}

\section{Conclusion}

In this paper, we have claimed and argued that software bots must produce explainable contributions. In order for them to seamlessly join the software development conversation, they have to be able to present and discuss their own  contributions: the intent, the design rationales, etc. 

In the context of a program repair bot such as Repairnator \cite{urli:hal-01691496,arXiv-1810.05806}, it means that the bot would  be able to reformulate the patch in natural language, to highlight the behavioral change with specific, well-chosen input values, to discuss why a particular patch is better than another one.

Beyond explainable contributions, we have hypothesized that software bots must have their own identify and their own character, so as to bring diversity in the development conversation. It may even be that the diversity of participants in a software development conversation is what makes it creative and fun.

\section*{Acknowledgments: } I would like to thank my research group for the fertile discussions on this topic, and esp. Matias Martinez and Khashayar Etemadi for feedback on a draft. This work was supported by the Wallenberg AI, Autonomous Systems and Software Program (WASP).

\balance
\bibliographystyle{abbrv}
\bibliography{biblio-erc.bib,biblio-software-repair.bib}

\end{document}